\documentclass[aps,prl,twocolumn,floatfix]{revtex4}
\usepackage{amsmath}
\usepackage{graphics}
\begin{document}
\title{First Principle Study of Electron Transport in Single-Walled 
Carbon Nanotubes of 2 to 22 nm in Length}
\author{Jun Jiang$^{1,2}$, Wei Lu$^2$ and Yi Luo$^1$\footnote{Corresponding author, Email: luo@theochem.kth.se}}
\affiliation{$^1$ Theoretical Chemistry, Royal Institute of Technology,
AlbaNova, S-106 91 Stockholm, Sweden}
\affiliation{$^2$ National Lab for Infrared Physics, Shanghai Institute of
Technical Physics, Chinese Academy of Sciences, China}
\date{\today}
\begin{abstract}
An elongation method based on {\it ab initio} quantum chemistry approaches is 
presented. It allows to study electronic structures and coherent 
electron transportation properties of single-walled carbon 
nanotubes (SWCNTs) up to 22nm in length using the hybrid density functional 
theory. The 22nm long SWCNT, consisting of more than ten thousands 
electrons, is the largest carbon nanotube that has ever been studied 
at such a sophisticated all-electron level. Interesting oscillating 
behaviour of the energy gap with respect to the length of the nanotube 
is revealed. The calculated current-voltage characteristics of SWCNTs 
are in excellent agreement with recent experimental results. It confirms
the experimental observation that a 15nm long SWCNT is still largely a 
ballistic transport device. The proposed elongation method opens up a new door 
for the first principle study of nano- and bio-electronics. 
\end{abstract}


\maketitle
\input epsf

Carbon nanotubes (CNTs) are probably the most studied nanomaterials in 
the last decade, owing to their great physical and chemical properties. 
Among many exciting applications, CNTs are believed to play an important 
role in the future electronics\cite{dai02surf,mceuen03IEEE,dressbook,dai04PNAS,dai04prl}.
This view is further enforced by an exciting recent development, namely the 
utilization of 10 to 50 nm long single-walled carbon nanotubes 
(SWCNTs)\cite{dai04PNAS,dai04prl}. It has been demonstrated that the electron 
transport in SWNTs with finite length less than 15nm immunes from the optical 
phonon scattering and thereby exhibits nearly ballistic behaviour at high 
biases\cite{dai04PNAS,dai04prl}. However, the finite-length SWNTs present a 
great challenge for the first principle theoretical modelling because of the 
involvement of vast number of electrons. Over the years, different approaches, 
such as semi-empirical\cite{sap98JPC,kwon94PRB}, {\it ab initio} tight-
binding\cite{aph00JPC,sankey89PRB,porezag95PRB,hansson00prb,yang04prb,guo00prl},
and "integrated" \cite{esp02PCPP,corc98JPC,corc98ACS,dapp99JMS} methods, have
been developed to describe the nano-sized systems, but all suffer from a 
relative low level of accuracy. For such large systems, the most accurate yet 
feasible approach is probably the density functional theory (DFT). In this 
letter, we present a straightforward elongation method in conjunction with 
the modern quantum chemical density functional theory calculations 
that allows to effectively treat very large nano-scale periodical systems 
without losing the accuracy. This method has applied for studying the 
electronic structures and the coherent electron transportation 
properties of SWCNTs of different lengths. The largest SWCNT reported 
here is about 22 nm long in length which consists of ten thousands 
of electrons and is described by fifteen thousands gaussian basis functions. 

Our elongation method is based on a simple fact that for a large enough 
finite periodic system, the interaction between different units in the 
middle of the system should be converged, and consequently those units in
the middle become identical. It is thus possible to elongate the initial 
system by adding the identical units in the middle of the system continuously. 
This can be easily done when the Hamiltonian of the system is describe 
in the site-representation. Obviously, the precondition for using the 
elongation method is to obtain an initial Hamiltonian in the 
site-representation with identical middle parts.  Fortunately this 
condition can be achieved routinely with the modern quantum 
chemistry programs. The elongation method should be as accurate as 
the quantum chemistry method used for the initial system. 
We will use the (5,5) metallic SWCNTs to demonstrate the 
performance of the elongation method as implemented in the 
QCME code\cite{qcme}. With the same code, the coherent electron transport 
in the SWCNTs is also calculated using the generalized quantum chemical 
Green's function method\cite{qcme,luo}. 

\begin{figure}
\vspace*{-0.0cm}
\begin{center}
\epsfxsize=240pt\epsfbox{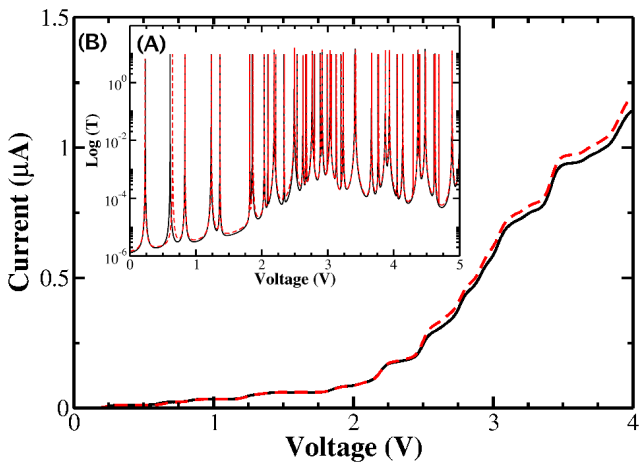} \vspace*{-0.0cm}
\hspace*{-0.0cm} \caption{(A) Transition probability (on a log
scale) spectrum above the Fermi level, and (B) I-V characteristics of
the CNT31 (310 carbons, 3.7nm in length) connected to two gold electrodes
obtained from Gaussian03 (solid lines) and elongation (dashed lines) methods.}

\label{comptightcnt310}
\end{center}
\vspace*{-0.0cm}
\end{figure}

We first calculate two (5,5) metallic SWCNTs with 190 (19 units,CNT19) and 310 (31 units, CNT31) carbons, respectively, at the hybrid density functional 
theory B3LYP level with STO-6G basis set using GAUSSIAN03 program\cite{gau03}. 
The both ends of the finite-sized tubes are terminated by hydrogen atoms. The 
calculated Hamiltonian of the CNT19 is used as the base for the 
elongation method to construct a Hamiltonian of the CNT31.  
The calculated highest occupied molecular orbital (HOMO) and lowest 
unoccupied molecular orbital (LUMO) energies from the
elongation method are found to be -2.000 and -1.538 eV, respectively, in 
excellent agreement with the values of -2.070 and -1.598 eV, respectively, 
obtained from GAUSSIAN03 program. The derivation for the LUMO-HOMO gap $E_g$ 
from these two different approaches is as small as 0.009 eV. The 
calculated transition probability spectra above the Fermi level 
for the CNT31 connected to two gold electrodes, as well as the 
corresponding current-voltage curves, from both methods, agree very well 
as demonstrated in Figure \ref{comptightcnt310}. It should be mentioned 
that the coupling coefficients between electrodes and CNTs is approximated by 
a simple Au-C diatomic interaction. The coupling between electrodes and CNTs  
is a complicated issue and will be
 discussed elsewhere. 

\begin{figure}
\vspace*{-0.0cm}
\begin{center}
\epsfxsize=240pt\epsfbox{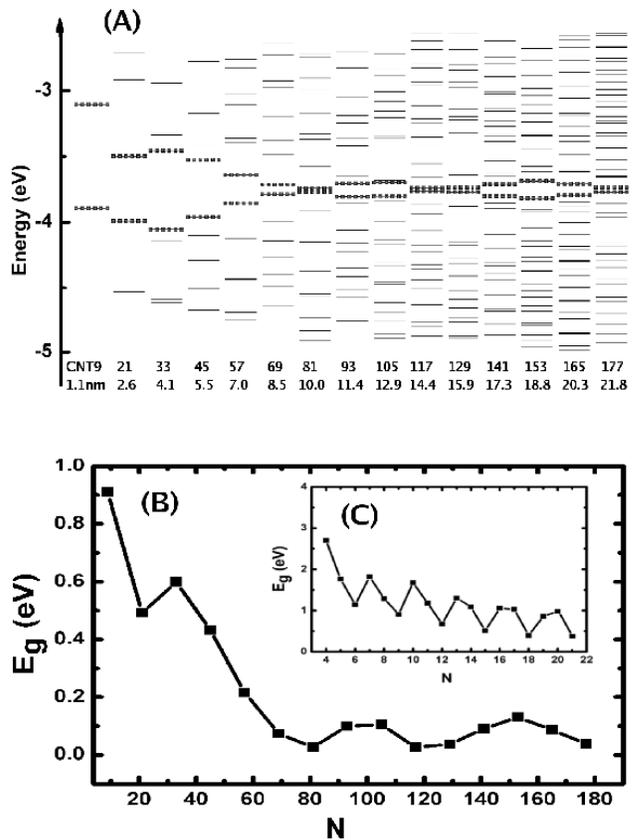} \vspace*{-0.0cm}
\hspace*{-0.0cm} \caption{Molecular orbitals of SWCNTs 
with $N=9+12 \times i$ ($i$=0, 1, ... ,14) units, corresponding to
1.1 nm to 21.8 nm in length (A). The results for the CNT9
and the CNT21 are calculated with GAUSSION03 while others are obtained
with the elongation method. Energy gap $E_g$ as a function of the units ($N$) 
is plotted for long-nano-scale (B) and short-nano-scale (C) SWNTs, 
respectively.} \label{cntallenergy}
\end{center}
\vspace*{-0.0cm}
\end{figure}

We are thus ready to explore the elongation method for even longer CNTs. 
To maintaining higher accuracy, we have used CNT21 (210 carbons, 2.6nm long) 
tube calculated at B3LYP/6-31G level with GAUSSIAN03 to construct a series of 
SWCNT systems with units of $n=9+12 \times i$ ($i$=2, 3,..., 14) 
whose length goes from 4.1 nm to 21.8 nm. In Figure \ref{cntallenergy} (A), 
the molecular orbitals (MOs) around the energy gap are displayed. 
As expected the density of states (DOS) gets higher for longer SWCNT. 
The most interesting observation is that the energy gap ($E_g$) 
oscillates periodically with the increase of SWCNT length. 
For short SWCNTs, it is known that the $E_g$ should oscillate
with a period of 3 units\cite{zhou04JACS,li02CPL,mat03ORG}. This behaviour 
is nicely reproduced by our calculations, see Figure \ref{cntallenergy}(C). 
The trend given by the short SWCNTs seems to indicate that the gap should
continuously decrease with the increase of the SWCNT length. However, 
our calculations present a very different picture. For instance, the energy 
gap of the CNT33 (0.598 eV) is already larger than that of the CNT21 (0.492 eV). 
This observation is further confirmed by Gaussian03 calculations 
with STO-6G basis set. The evolution of energy gap with respect to 
the tube length is clearly shown in Figure \ref{cntallenergy}(B). It is 
noticed that the width of the oscillation peak covers 36 units from the CNT81 
to the CNT117, while the one after it is about 60 units from the CNT117 to 
the CNT177. It seems to imply that the oscillation of the energy gap 
with respect to the length of the tube will be slowly faded out. 
Furthermore, discrete molecular orbital distribution is found for 
all the tubes under investigation, indicating that these tubes can be
quantized electronic devices. One can at least conclude
that a (5,5) SWCNT with length of 22 nm is still far from a real bulk material.  

\begin{figure}
\vspace*{-0.0cm}
\begin{center}
\epsfxsize=240pt\epsfbox{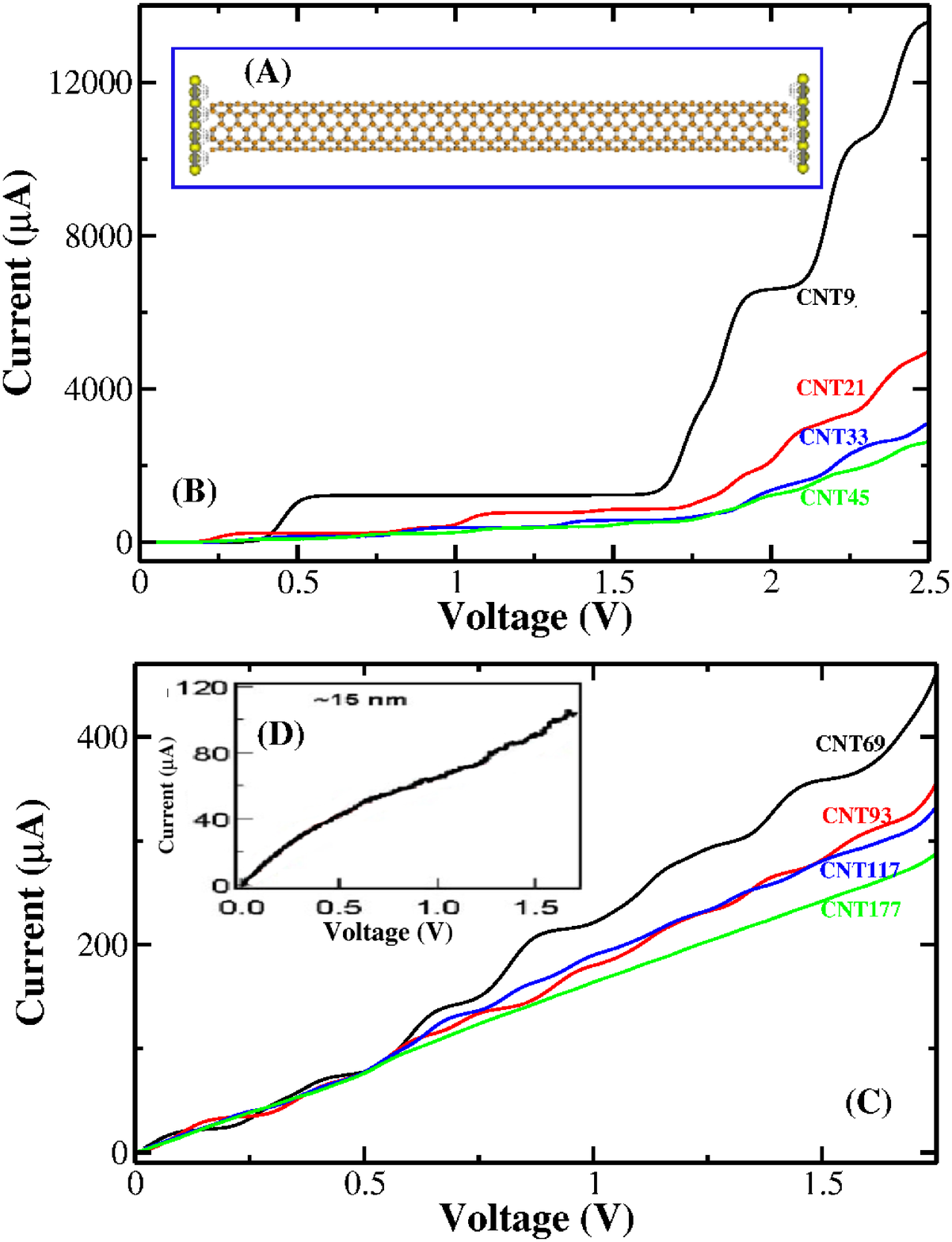} \vspace*{-0.0cm}
\hspace*{-0.0cm} \caption{(A): Schematic draw of a SWCNT device with gold
electrodes; (B) Calculated current-voltage characteristics for the CNT9 (1.1nm), 
the CNT21 (2.6nm), the CNT33 (4.1nm), and the CNT45 (5.5nm); (C) Calculated
current-voltage characteristics for the CNT69 (8.5nm), the CNT93 (11.4nm), 
the CNT117 (14.4nm), and the CNT177 (21.8nm); and (D) Experimental I-V curve 
for a 15nm SWCNT\cite{dai04PNAS}.} \label{allcntcur}
\end{center}
\vspace*{-0.0cm}
\end{figure}

By using the QCME code, we are able to calculate the I-V characteristics of
the tubes with different lengths. The electrode structure shown in 
Figure \ref{allcntcur}(A) is the same as that reported in the 
experimental work of Javey et al\cite{dai04PNAS}. The I-V curves of 
short tubes up to the CNT45 (5.5nm) show stair-like profile in which 
each step represents an opening of a new conducting channel. 
These small devices are all semi-conductor-like, i.e. the electron stars 
to flow only after adding certain external bias. When the length of the tube
gets longer, the DOS of the system becomes denser, the steps in the I-V 
curve smear out, and the device shows a metallic behaviour. We have compared our
calculations with the experimental result\cite{dai04PNAS}. As indicated 
in Figure \ref{allcntcur}(D), our calculated I-V curve for a 14.4nm tube (CNT117) 
agrees surprisingly well with the experimental result of a 15nm long tube for 
both the shape and the magnitude. It thus confirms the experimental observation
that a device with a 15nm long tube behaves like the ballistic
conductor\cite{dai04PNAS}.

\begin{figure}
\vspace*{-0.0cm}
\begin{center}
\epsfxsize=240pt\epsfbox{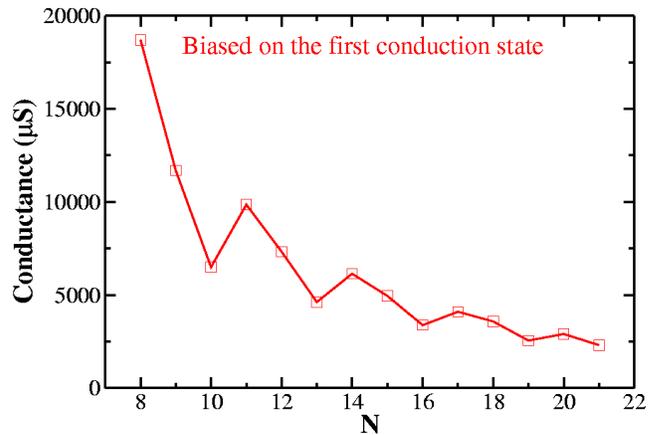} \vspace*{-0.0cm}
\hspace*{-0.0cm} \caption{The conductance for SWCNTs with 
units $N=$8, 9, ..., 21 biased at their first conduction state.} \label{cntperiod}
\end{center}
\vspace*{-0.0cm}
\end{figure}

It might be anticipated that the oscillation of the energy gap could result
in a similar behaviour for the electron conductivity. Indeed, such behaviour 
is clearly observed for the conductance of different SWCNTs, see 
Figure \ref{cntperiod}. The period for the oscillation is also the same as that 
for the energy gap, i.e. 3 units. However, the phase of these two oscillations 
is slightly shifted. The tubes with $3n+1$ units have the lowest conductance, 
while the tubes with 3n units give the smallest energy gap.

\begin{figure}
\vspace*{-0.0cm}
\begin{center}
\epsfxsize=240pt\epsfbox{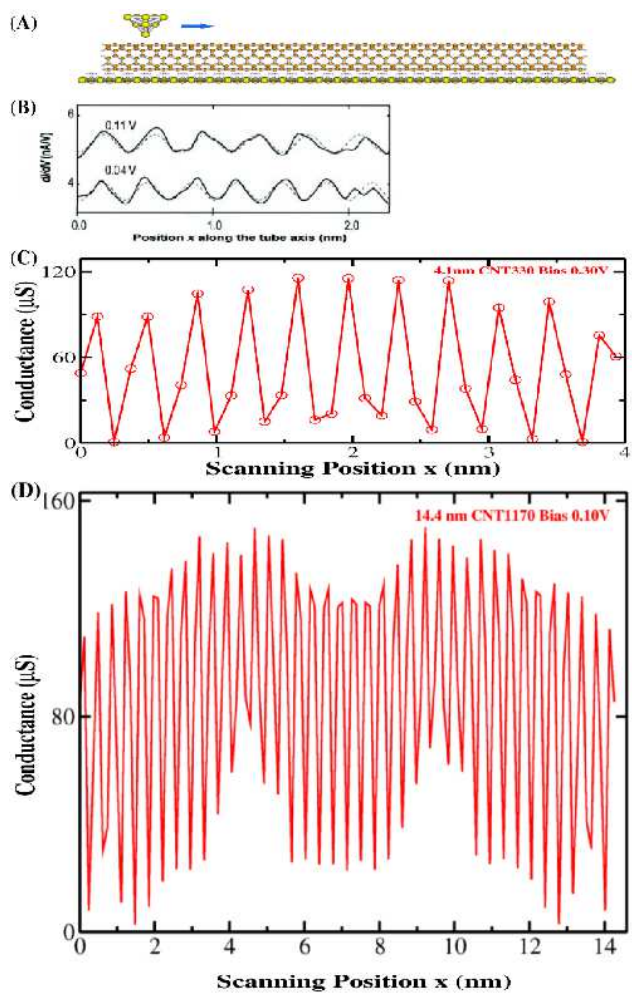} \vspace*{-0.0cm}
\hspace*{-0.0cm} \caption{Schematic draw of 
STM scanning experiment (A) and the corresponding conductance measurement B) 
for a 30nm SWCNT\cite{venema99sci}. Calculated position dependent conductance 
for the CNT33 (4.1nm) at external bias 0.3V and for the CNT117 (14.4nm) at 0.1V.} \label{cntperiod1}
\end{center}
\vspace*{-0.0cm}
\end{figure}

Another interesting quantum effect related to the SWCNT is the periodical 
distribution of its molecular orbitals along the tube axis.
Venema et al.\cite{venema99sci} carried out the position-scanning I-V measurements for a finite-length metallic SWCNT. The experimental setup is 
schematically recaptured in Figure \ref{cntperiod1} together with the 
experimentally measured result. In this experiment, a STM tip scanned 
through the top of a 30nm metallic SWCNT deposited on a gold surface to 
measure the current at each positions. It was found that the conductance 
of this tube shows a periodic behaviour with a period of about 0.4 nm, 
roughly covering 3 layers. With the elongation method, we have simulated 
the STM experiment for two tubes, the CNT33(4.1nm) and the CNT117 (14.4nm) 
at external biases of 0.3V and 0.1 V respectively. The results are also shown 
in Figure \ref{cntperiod1}. As one can see, the oscillating behaviour is 
well reproduced, which shows a period of 3 layers. It should be stressed 
that our calculations describe only the coherent electron transport, while 
for a 30nm long SWCNT phonon coupling can become important\cite{dai04PNAS} 
which might explain why the periodicity of the experimental result is less 
perfect. Furthermore, there is a very interesting wave-packet-like 
behaviour in the position depended conductance of the CNT117. It seems to 
suggest that an electron standing-wave can emerge in SWCNT longer than 15 nm.

In summary, we have proposed an elongation method that is capable of calculating
electronic structures and electron transportation properties of periodic nano-
sized systems at the hybrid density functional theory level. With this method, 
we have studied the length dependence of the energy gap and the conductivity 
of the finite-length single-walled carbon nanotubes (SWCNTs) ranging 
from 1nm to 22nm. The calculated results agree very well with the full 
quantum chemical calculations and exiting experimental results. A new 
interesting oscillating behaviour for the energy gap of the finite-length 
SWCNTs is revealed. It is worth to mention that our elongation has also 
successfully applied to 50nm long poly(para-phenylene ethynylene)s and 
60 pairs poly(G)-poly(C) DNA. The calculated I-V characteristics of these 
systems are also in very good agreement with the corresponding experiments. 
The details of these studies will be published elsewhere.  

\vskip 0.5truecm
\centerline{\bf Acknowledgments}
\vskip 0.2truecm
This work was supported by the Swedish Research Council (VR), 
the Carl Trygger Foundation (CTS), 
and Chinese state key program for basic research(2004CB619004).

\end{document}